\documentclass[conference]{IEEEtran}
\IEEEoverridecommandlockouts
\usepackage{cite}
\usepackage{amsmath,amssymb,amsfonts}
\usepackage{algorithmic}
\usepackage{graphicx}
\usepackage{textcomp}
\usepackage{xcolor}
\usepackage{tabularx}
\usepackage{adjustbox}
\usepackage{enumitem}
\usepackage{siunitx}
\usepackage{comment}
\usepackage{xcolor, soul}
\sethlcolor{yellow}
\usepackage[compact]{titlesec}
\usepackage{url}
\titlespacing{\subsection}{0pt}{*0}{*0}
\usepackage[
singlelinecheck=false 
]{caption}
\usepackage{xcolor}
\usepackage{caption}
\usepackage{tabularray}
\usepackage{amsmath}
\usepackage{listings}
\UseTblrLibrary{booktabs}
\def\BibTeX{{\rm B\kern-.05em{\sc i\kern-.025em b}\kern-.08em
    T\kern-.1667em\lower.7ex\hbox{E}\kern-.125emX}}

\makeatletter
 \let\old@ps@headings\ps@headings
 \let\old@ps@IEEEtitlepagestyle\ps@IEEEtitlepagestyle
 \def\confheader#1{%
 \def\ps@headings{%
 \old@ps@headings%
 \def\@oddhead{\strut\hfill#1\hfill\strut}%
 \def\@evenhead{\strut\hfill#1\hfill\strut}%
 }%
 \def\ps@IEEEtitlepagestyle{%
 \old@ps@IEEEtitlepagestyle%
 \def\@oddhead{\strut\hfill#1\hfill\strut}%
 \def\@evenhead{\strut\hfill#1\hfill\strut}%
 }%
 \ps@headings%
 }
\makeatother
\usepackage{hyperref}
\usepackage[T1]{fontenc}
\usepackage{tikz}
\usetikzlibrary{arrows, calc, decorations.markings, positioning}

\usepackage{adjustbox}
\usepackage{tabularx}
\usepackage{todonotes}
\usepackage{array, multirow, boldline}
\usepackage{tcolorbox}

\usepackage{forest}
\usetikzlibrary{shadows}

\usepackage{color, colortbl}
\definecolor{Gray}{gray}{0.8}
\definecolor{LightGray}{gray}{0.95}

\newcounter{rowcntr}[table]
\renewcommand{\therowcntr}{\thetable.\arabic{rowcntr}}
\newcolumntype{N}{>{\refstepcounter{rowcntr}\therowcntr}c}
\begin{document}

\title{A Lightweight Edge-CNN-Transformer Model for Detecting Coordinated Cyber and Digital Twin Attacks in Cooperative Smart Farming}


\author{\IEEEauthorblockN{Lopamudra Praharaj\IEEEauthorrefmark{1}, Deepti Gupta\IEEEauthorrefmark{2}, and Maanak Gupta\IEEEauthorrefmark{3}}
\IEEEauthorblockA{\IEEEauthorrefmark{1}\IEEEauthorrefmark{3}{Dept. of Computer Science},
{Tennessee Tech University},
Cookeville, Tennessee 38505, USA \\\IEEEauthorrefmark{2}{Dept. of Computer Information Systems}, Texas A\&M University - Central Texas, TX, USA\\}
\IEEEauthorrefmark{1}lpraharaj42@tntech.edu, 
\IEEEauthorrefmark{2}d.gupta@tamuct.edu,
\IEEEauthorrefmark{3}mgupta@tntech.edu}
\maketitle

\begin{abstract}
The agriculture sector is increasingly adopting innovative technologies to meet the growing food demands of the global population. To optimize resource utilization and minimize crop losses, farmers are joining cooperatives to share their data and resources among member farms. However, while farmers benefit from this data sharing and interconnection, it exposes them to cybersecurity threats and privacy concerns. A cyberattack on one farm can have widespread consequences, affecting the targeted farm as well as all member farms within a cooperative. For instance, a de-authentication attack prevents sensors from connecting to the network, obstructing the farming application from receiving real-time data. This disruption obstructs decision-making for all member farms that rely on this data from an attacked farm. Further, farmers have adopted digital twin (DT) technology that facilitates a virtual farm replica that encompasses vital aspects of farming, such as crop cultivation, soil composition and weather conditions. However, it's critical to acknowledge that attackers can target these digital twins (DTs), potentially disrupting real physical farm operations.

In this research, we address existing gaps by proposing a novel and secure architecture for Cooperative Smart Farming (CSF). First, we highlight the role of edge-based DTs in enhancing the efficiency and resilience of agricultural operations. To validate this, we develop a test environment for CSF, implementing various cyberattacks on both the DTs and their physical counterparts using different attack vectors. We collect two smart farming network datasets to identify potential threats. After identifying these threats, we focus on preventing the transmission of malicious data from compromised farms to the central cloud server. To achieve this, we propose a CNN-Transformer-based network anomaly detection model, specifically designed for deployment at the edge. As a proof of concept, we implement this model and evaluate its performance by varying the number of encoder layers. Additionally, we apply Post-Quantization to compress the model and demonstrate the impact of compression on its performance in edge environments. Finally, we compare the model's performance with traditional machine learning approaches to assess its overall effectiveness.

\end{abstract}

\section{Introduction}
Recent studies~\cite{roser2013world}, \cite{godfray2010food} present a substantial increase in the global population, expected to reach 9.7 billion by 2050. In response to this growth, it's vital to utilize advanced technologies like the Internet of Things (IoT), Machine Learning (ML), Cloud Computing, and autonomous systems to manage the balance between food supply and demand effectively.

Incorporating connected sensors with cloud and edge services in agriculture enables farmers to make automated decisions \cite{chukkapalli2020ontologies}. As individual smart farms adopt more data-driven technologies, their capital investment and maintenance costs increase. Although many farmers are eager to embrace modern technologies, the cost and technical expertise required often present significant obstacles for small-scale farmers. 
Thus, through a model called Cooperative Smart Farming (CSF), small-scale farmers can reap the benefits of deploying sensors and AI-based solutions. These cooperatives are formal enterprises that are financed, controlled, and owned by member farms for mutual benefits, including resource sharing, financial assistance, access to large machinery, efficient labor during peak fieldwork periods, production transportation, price negotiation, equipment repair and maintenance, and data sharing, among others \cite{chukkapalli2020ontologies,chukkapalli2021privacy}.

Cyberattacks on agricultural systems are no longer a distant threat; they are a present and growing reality. As per the Federal Bureau of Investigation (FBI) \cite{cyberthreats}, agricultural cooperatives are threatened by well-timed ransomware attacks. In early 2021, a ransomware assault targeted meat producer JBS and two-grain purchasers in the United States during the harvest season. In September 2021, a ransomware attack targeted an Iowa-based cooperative, demanding a ransom of 5.9 million dollars. Additionally, in February 2022, a company providing feed milling and agricultural services reported unauthorized access attempts on some of its systems \cite{cyberthreats,realcyberattack1}.

CSF faces a greater risk of cyberattacks than individual farms due to their interconnected networks and extensive data sharing among member farms.
If one smart farm within the cooperative falls victim to an attack, it also impacts other farms. For instance, let's envision a scenario where \textit{Smart Farm A} installs temperature, humidity, rainfall, and soil moisture sensors in its fields. {\textit{Smart Farm B}, a cooperative member, relies on the soil moisture sensor data from \textit{Smart Farm A} to determine when to plant crops in its fields. However, due to a cyber attack on \textit{Smart Farm A}, the soil moisture sensor reading is manipulated from 45 to 350. Consequently, \textit{Smart Farm B} and other smart farms utilizing this data believe the soil moisture level is moist when in a cooperative environment; in reality, it is dry. As a result, the attacked farm and other smart farms within the cooperative make incorrect farming decisions based on this erroneous data. The example above demonstrates a "Coordinated cyberattack" on farms within a cooperative environment. In this context, Coordination refers to simultaneously targeting multiple farms. Even if a malicious attacker can attack a specific farm, other farms can still suffer cascading effects within the cooperative environment.


Recent research \cite{angin2020agrilora, alves2019digital, purcell2023digital} focus on the necessity of farmland Digital Twins (DTs), which allow for monitoring the status of farmlands and predicting their physical counterparts in near real-time through their virtual representation and taking appropriate actions recommended by intelligent processing of the gathered data. 
While much of the existing research on DT in agriculture focuses on cloud-based implementations, recent studies highlight the advantages of deploying DTs at the edge \cite{protner2021edge}. By leveraging edge computing, farmers can access DT data locally, optimising resources such as water, fertilizers, and pesticides more efficiently. Each farm can maintain its own DT model on its local edge, while other member farms can access and utilize DT data as needed, as per the cooperative agreement.
 However, these DTs are also susceptible to cyberattacks \cite{suhail2022security,karaarslan2021digital}, where attackers may compromise them by injecting malicious code to corrupt the DT model at the edge. 

Thus, cyberattackers can target both DTs and their physical counterparts using different attack vectors. To detect and prevent cyberattacks, edge-based network anomaly detection has proven to be an effective approach that utilizes ML techniques to develop models that distinguish between normal and anomaly behaviour based on observed features \cite{dong2021network, mahmoudzadeh2020spatial}. This approach ensures that if an anomaly detector identifies a smart farm as compromised,  an alert mechanism notifies the affected farm, prompting it to isolate from the cooperative network and thereby preventing the spread of cyberattacks within the CSF.

Recently, transformers and their variants, which leverage the self-attention mechanism, have demonstrated remarkable success in Natural Language Processing (NLP) tasks such as text classification, dialogue recognition, and machine translation \cite{gillioz2020overview, kalyan2021ammus}. Additionally, due to their ability to manage long data sequences, researchers have increasingly utilized transformers to detect anomalies and intrusion detection across various scenarios \cite{liu2023intrusion, wu2022rtids}. 

In this paper, we implement a CNN-Transformer-based network anomaly detector to detect and mitigate cyberattacks that target DTs and their physical counterparts on the edge layer. The CNN component is responsible for extracting spatial features from the data, such as local patterns and relationships, while the Transformer component leverages its self-attention mechanism to capture long-range dependencies and contextual information in the network data. The softmax function is applied to classify the network attacks. The proposed model combines the advantages of CNN and Transformer and has excellent anomaly detection performance with the increasing number of encoding layers. However, with an increasing number of encoding layers, it requires significant memory, making deployment on edge devices challenging \cite{jena2024unified, kulkarni2021ai}, so we compress the model using the Post-Quantization technique \cite{sharmila2023qae, lu2023hierarchical,jena2024unified} and showcase the advantage of model compression at the edge. The main contributions of the papers are as follows:
\begin{enumerate}
    \item We identify the challenges inherent in CSF, outline the layers that enhance the farming operations and showcase the importance of the edge layer, particularly with the integration of Digital Twin (DT) technology at the edge in CSF.
    \item We develop a testbed environment using an Arduino ESP8266 NodeMCU board and various sensors commonly used in smart farming. Additionally, we implement all the layers (physical, edge, and cloud) as per our proposed architecture using AWS and Azure technologies. We also implement a digital twin at the edge using Azure Digital Twin and subject it to attacks. 
    \item  We simulate two smart farming datasets relevant to CSF scenarios, where each smart farm was subjected to various network attacks using different sensors employed in smart farming. 
    
    \item We propose a CNN-Transformer-based network anomaly detection model to detect and mitigate cyberattacks in CSF and evaluate its performance with two smart farming datasets by varying the number of encoder layers. To facilitate effective edge deployment, we compress the model using Post-Quantization techniques and compare its performance with other machine learning models.
\end{enumerate}
The remainder of this paper is organized as follows. Section \ref{sec:rel} discusses relevant literature. Section \ref{sec:architecture} addresses Proposed Edge Enabled CSF architecture. Section \ref{sec:testbed} defines the steps in building the realistic test bed in the CSF environment and Digital Twin using AWS and Azure Technologies. Section \ref{sec:cyberattacks} defines the implementation of various cyber attacks at the Network, DT and collection of two smart farming datasets. Section \ref{sec:networkarchitecture} provides the details of Edge-based network anomaly detection architecture and the CNN-Transfomer Model. Section \ref{sec:analysis} discusses the implementation of the model and evaluation of compression technology on our networking datasets. The conclusion and future works are presented in Section \ref{sec:conclusion}.
\section{Related Work}
\label{sec:rel}
\begin{figure*}[!t]
\centering
\includegraphics[width=0.9\textwidth]{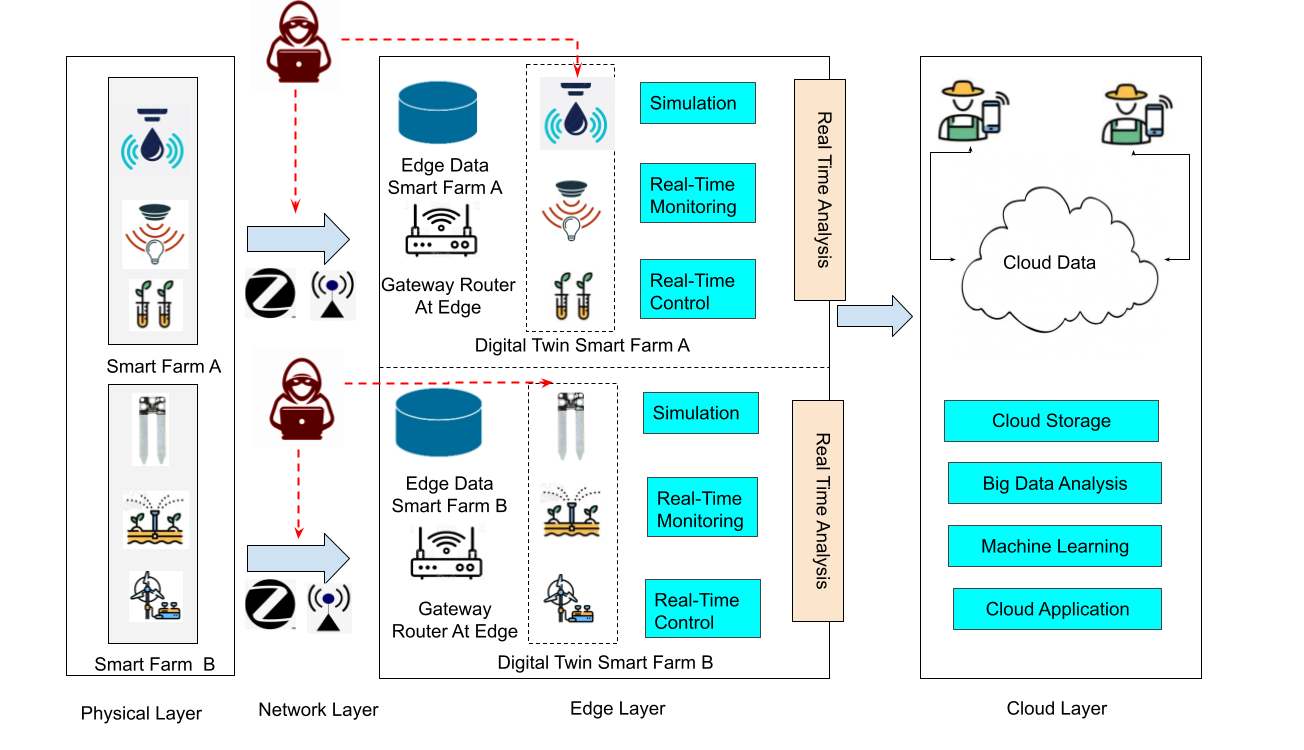}
\caption{Layer Based Cooperative Smart Farming Architecture.}
\label{fig:fig1}
\end{figure*}
\subsection{Cooperative Smart Farming}
Cooperatives represent formal enterprises structured and managed by their members to achieve mutual benefits. Through cooperative agreements, farmers collaborate by sharing valuable resources such as sensor data, drone imagery, and agricultural equipment. SSL Chakkrupali's \cite{chukkapalli2020ontologies} work outlines the development of a cohesive ecosystem, detailing various components like on-farm sensors, drones, and a range of farm machinery. Furthermore, the paper explored diverse AI applications tailored for cooperative ecosystems.  

In the subsequent paper~\cite{chukkapalli2021privacy}, the authors utilized smart farm ontology to perform data transformation and thus populated the knowledge graph by adding the transformed data to cooperative ontology. In \cite{gupta2020game, gupta2020learner, gupta2022game}, the authors identified the issue of unfair collaboration among farms in CSF and propose a fair strategy to force farms to cooperate in CSF to build ML model. In \cite{praharaj2023hierarchical}, the authors harnessed the advantages of DT technology within a CSF context and introduced a hierarchical federated transfer learning approach based on CNN-LSTM as an effective AD model. This model aids in the early detection of threats and safeguards against zero-day attacks in the CSF environment.
\subsection{Anomaly  Detection  in  Smart Farming Using Deep Learning Techniques}
Researchers have introduced a deep learning-based network anomaly detection strategy into smart agriculture to address emerging challenges based on excellent results in smart IoT systems. In \cite{mahmoudzadeh2020spatial}, a robust intrusion detection system for DDoS attacks in smart agriculture was developed. A CNN algorithm was combined with the Bi-GRU model to both detect and classify intrusions. The attention mechanism within the Bi-GRU model identifies the most critical features needed to recognize DDoS attacks effectively. Ferrag et al. \cite{ferrag2021deep} explored using CNNs, deep neural networks, and recurrent neural networks, specifically evaluating the effectiveness of deep learning and machine learning methods for cybersecurity in Agriculture 4.0. The study assessed the performance of these models using real traffic datasets, CICDDoS2019 and TON IoT, for both binary and multi-class classifications. Cheng et al. \cite{cheng2022anomaly} proposed a GAN-based anomaly detection algorithm for multi-dimensional time series data generated by the smart agricultural Internet of Things. The model learned the distribution patterns of normal data using the GAN architecture and utilized reconstruction techniques to identify anomalies. Di Mauro et al. \cite {di2020experimental} critically compared approaches with a common paradigm, including intrusion detection methods such as weightless neural networks. The authors evaluated the model using the CIC-IDS-2017/2018 dataset, focusing on binary and multi-class classification.
Dong, S. et al. \cite{dong2021network} introduced a network anomaly detection model based on semi-supervised deep reinforcement learning. The model utilized a semi-supervised double deep Q-network (SSDDQN) to optimize the detection of abnormal network traffic, leveraging the double deep Q-network approach.

\section{Proposed Edge Enabled Cooperative Smart Farming Architecture with Digital Twins}
\label{sec:architecture}

\begin{figure*}[!t]
\centering
\includegraphics[width=0.9\textwidth]{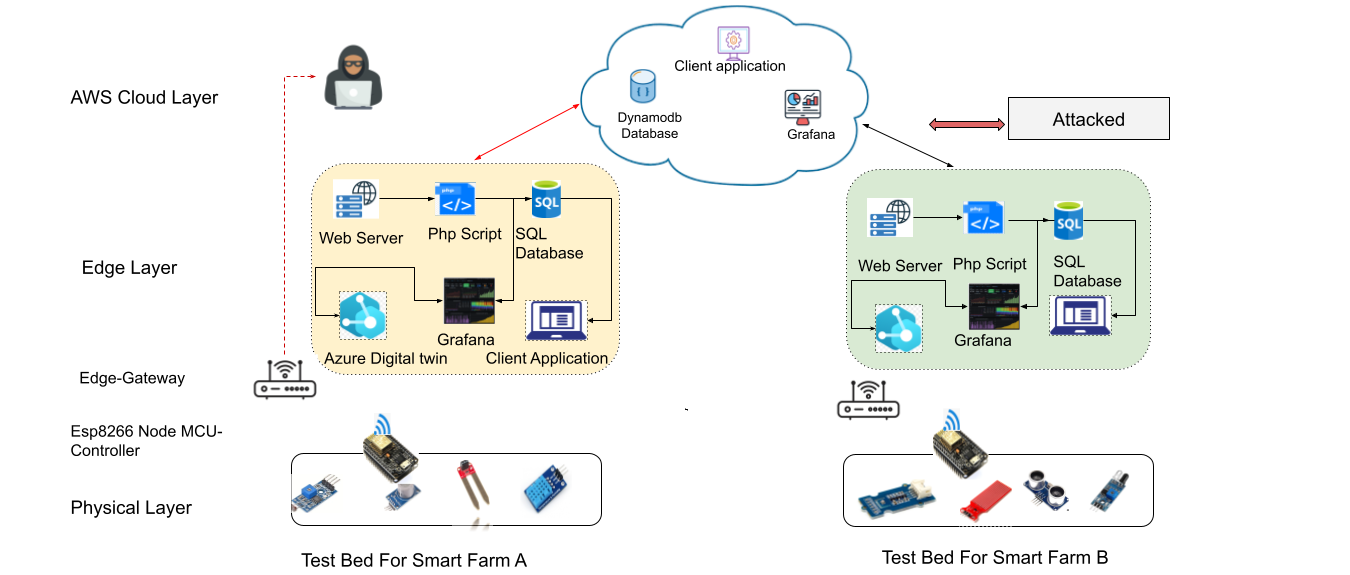}
\caption{Test Bed Environment for Cooperative Smart Farming}
\label{fig:fig2}
\end{figure*}
CSF involves multiple farms collaborating to share resources, data, and technology to optimize agricultural practices. 
In this section, we present the distinct layers of the CSF, illustrating how these collaborative efforts are structured and interconnected. Categorized by their respective functionalities, the layers are classified into three main divisions: The physical layer, the edge layer, and the cloud layer. The section will concisely overview the role and purpose of each layer.
\subsection{Physical Layer}
As shown in Fig.~\ref{fig:fig1}, the physical layer encompasses sensors and devices with real-time data sensing derived from the collected information. These devices, ranging from drones and automated tractors to various sensors and actuators, play a crucial role in immediately acquiring data. Within a cooperative network of smart farms, sensors are strategically deployed to monitor fields based on specific requirements. The data collected at the physical layer is subsequently transmitted to the corresponding edge for more in-depth analysis. As shown in Fig. \ref{fig:fig1}, each smart farm in a cooperative is equipped with various physical devices. For instance, in \textit{Smart Farm A}, the soil moisture sensor collects information and sends it to the edge layer at \textit{Smart Farm A} for further processing and evaluation.
\subsection{Edge Layer}
Data from the physical layer is transmitted to the edge layer using wireless network devices like gateway routers. The decision to route data to the edge instead of directly sending it to the cloud offers numerous advantages. Edge computing allows for data processing nearer to where it is generated, reducing the need for extensive data transfers to a central cloud server. This feature enhances bandwidth efficiency and ensures critical data is readily available for local decision-making \cite{sha2017edgesec, portilla2019extreme}. By processing data near its source, edge computing significantly reduces latency. This feature is particularly vital in smart farming applications where rapid responses to field conditions are essential, making low-latency operations imperative. Edge computing involves distributing data processing and storage across local devices, enhancing security and reducing the risk of exposing sensitive agricultural data.
To further advance the capabilities of smart farming, we present a high-precision, extensible IoT-based farm field DT at the edge (Fig. \ref{fig:fig1}). At its core, the DT is a virtual replica of the physical world, enabling comprehensive insights and actionable data for optimal farm management \cite{purcell2023digital,alves2019digital}. A DT continuously receives data from its physical counterpart to provide an up-to-date virtual model, and the virtual model can also provide feedback to the same communication channel. The DT has significant benefits when deployed at the edge; it can reduce latency and enable real-time decision-making. For example, the Edge-based DT can immediately adjust the irrigation schedule or the amount of water delivered to that area without waiting for data from a central cloud server. It also offers several advantages when coming to the security of smart farming. Those are behavioural analysis, vulnerability assessment, historical data analysis, real-time alerts, and simulation of attack scenarios \cite{alves2019digital, angin2020agrilora, nasirahmadi2022toward}.
\subsection{Cloud Layer}
The sensor data produced by both smart farms is stored in cloud storage for later analysis (Fig. \ref{fig:fig1}). Access to this data is accelerated through a shared application in the cloud, allowing the smart farms to collaborate and exchange information in CSF. However, if the data is altered or inaccessible due to a cyber attack, it affects both the smart farms. 

\section{Test Bed Environment}
\label{sec:testbed}

\begin{table*}[t]
\centering
  \caption{List of all Cyber-attacks and its effects in Cooperative Smart Farming TestBed}
  \label{tab:attack-summary}
  
  \begin{tabular}{ccccc}
    \hline
    \rowcolor{Gray}
    \multicolumn{1}{p{3cm}}{\centering Attack Category}
    & \multicolumn{1}{p{3cm}}{\centering Attack Type}
    & \multicolumn{1}{p{4cm}}{\centering Vulnerabilities due to the attack}
    & \multicolumn{1}{p{4cm}}{\centering Effects on the Cooperative Smart Farms}
    & \multicolumn{1}{p{2cm}}{\centering Tools}

    \\
    \multicolumn{1}{p{3cm}}{\centering DoS Attack}
    & \multicolumn{1}{p{3cm}}{\centering de-authentication attack}
    & \multicolumn{1}{p{4cm}}{\centering Render the victim's edge server incapable of processing requests}
    & \multicolumn{1}{p{4cm}}{\centering Smart Farms are unable to access up-to-date data because of the unavailability of the server}
    & \multicolumn{1}{p{2cm}}{\centering Airplay-ng}
    
    \\
    
    \rowcolor{LightGray}
    
    & \multicolumn{1}{p{3cm}}{\centering TCP SYN Flood Attack}
    & \multicolumn{1}{p{4cm}}{\centering The attacker sends a barrage of SYN requests to a target server }
    &

     & \multicolumn{1}{p{2cm}}{\centering Hping3}
    
    \\
    \\
    \multicolumn{1}{p{3cm}}{\centering 
Man in the Middle Attack
}
    & \multicolumn{1}{p{3cm}}{\centering ARP Poisoning Attack 
}
    & \multicolumn{1}{p{4cm}}{\centering 
Associating the MAC address of the attacker with the IP address of the edge server can intercept network traffic}
    & \multicolumn{1}{p{4cm}}{\centering  Visualize and intercept data belonging to the smart farm, leading to damage to the cooperative's reputation.}
    & \multicolumn{1}{p{2cm}}{\centering Ettercap Tool}
    
    \\
    
    \rowcolor{LightGray}
    
    & \multicolumn{1}{p{3cm}}{\centering DNS Spoofing Attack}
    & \multicolumn{1}{p{4cm}}{\centering Intercepting communication between an ESP8266 device and the edge DNS server.}
    & \multicolumn{1}{p{4cm}}{\centering The farmer is forced to redirect communication to a fake server instead of the legitimate one, causing service disruptions. }
& \multicolumn{1}{p{2cm}}{\centering Ettercap tool}
    \\

    & \multicolumn{1}{p{3cm}}{\centering Evil Twin Attack}
    & \multicolumn{1}{p{4cm}}{\centering 
Create a fraudulent access point mimicking the router to lure victims into connecting. Then, spoof all packets passing through the ESP8266 device to intercept and manipulate communication.
}
    & \multicolumn{1}{p{4cm}}{\centering Leads to unauthorized access to sensitive data, disruption of communication networks, and potential financial losses due to compromised systems.}
    & \multicolumn{1}{p{2cm}}{\centering Airdeggon Tool}
    \\

    \multicolumn{1}{p{3cm}}{\centering Injection Attack}
     
    & \multicolumn{1}{p{3cm}}{\centering SQL Injection Attack}
    & \multicolumn{1}{p{4cm}}{\centering Inject the false data to the SQL server using the false query.}
    & \multicolumn{1}{p{4cm}}{\centering Manipulated data stored in the database server have impacted the farmer's decision-making process}
    & \multicolumn{1}{p{2cm}}{\centering SQL map Tool}
    
   \\

      \rowcolor{LightGray}
    \multicolumn{1}{p{3cm}}{\centering 
Information Gathering/ Reconnaissance  Attack
}
    & \multicolumn{1}{p{3cm}}{\centering Port Scanning, OS Fingerprinting}
    & \multicolumn{1}{p{4cm}}{\centering Discover the vulnerability in the smart farm Network}
    & \multicolumn{1}{p{4cm}}{\centering The attacker gathers all the information, such as weak points of the network, devices and the edge server.}
    & \multicolumn{1}{p{2cm}}{\centering Nmap and Xprobe2}
  \end{tabular}
  
\end{table*}
In our study, to illustrate the impact of cyberattacks on the smart farms within CSF environment, we develop a realistic testbed that closely replicates a real-world IoT environment for the CSF. We demonstrate cyberattacks against this testbed. (Fig. \ref{fig:fig2}) to support our proposed work. 

To demonstrate coordinated cyber attacks in smart farming, we conduct simulations involving two distinct smart farms configurations, each equipped with varying arrangements of sensors and edge devices. The testbed consists of three interconnected layers: physical layer, edge layer, and cloud layer.
The following are the primary microcontroller components utilized for the development of the test bed set up for \textit{Smart Farm A}: (1) ESP8266 (NodeMCU), (2) Temp and humidity sensor, (3) Gas sensor, (4) Light sensor, (5) Soil Moisture sensor.
These sensors are specifically chosen for their applicability in monitoring smart farms. The Light Sensor ensures plants receive sufficient light, the Air Gas Sensor detects potentially harmful gases, and the Soil Moisture Sensor helps maintain appropriate soil moisture levels. On the other hand, the temperature and humidity sensor monitors temperature and humidity, providing vital data for creating optimal growing conditions for indoor plants.

C++ programming is executed through the Arduino IDE to facilitate communication between Esp8266 NodeMCU and all sensors.
Additionally, it utilizes the ESP8266 NodeMCU to establish WiFi network communication between the sensors and the edge device (Fig. \ref{fig:fig2}).
Subsequently, the ESP8266 interacts with the MySQL server through the following steps:
\begin{itemize}

\item \textbf{Step 1}: The esp8266 incorporates sensor data into an HTTP request, sending the request to the web browser.
\item \textbf{Step 2}: A web server executes a PHP script responsible for handling the request originating from ESP8266.
\item \textbf{Step 3}: The PHP script extracts the data from the HTTP request and inserts it into the MySQL Database.
\item \textbf{Step 4}: Following the insertion of data, we leveraged AWS Grafana \cite{grafana} to enhance data visualization.
\end{itemize}
We install the MySQL server, Web server, PHP script, and AWS Grafana on the local edge.
In creating a real environment for CSF, we configure diverse sensors and introduce an additional edge device for \textit{Smart Farm B}.
Data from both smart farms can be uploaded to the AWS cloud layer. We opt for AWS services due to their cost-effectiveness and faster cloud capabilities. Specifically, we utilize the AWS NoSQL database to store the data in the cloud and a single cloud application where both smart farms can access sensor data, subject to authorization policies. Moreover, smart farms can visualize sensor data using AWS-managed Grafana services \cite{awsgrafana}. This setup mirrors a real-world farm scenario, contributing to emulating CSF environment.

\subsection{Digital Twin Environment}
Among the numerous companies offering DT services (Siemens MindSphere Edge, IBM Edge Application Manager, Bosch IoT Edge, Cisco DT) at the edge, we opt for Azure DT services \cite{azuretwin} due to their ease in modelling and creating digital representations of connected environments facilitated by an open modelling language. To establish DT at the edge, we set up the Azure digital twin instance and configure the appropriate roles. We execute "az login" in the command prompt to authenticate the process. Subsequently, we download the zip file "digitals-twin-explorer.zip" and install it on the local host \cite{azuretwin}. The Azure DT Explorer will then operate on the local host, typically on port number 5000 by default. Next, we create the DT model utilizing the DTDL language. We simulate the sensors to replicate the smart farm models, mirroring their physical counterparts. 
To establish the bridge between the physical layer and Azure DT, we develop a client application in Python. This application connects to the SQL server and the Azure DT instance. It efficiently retrieves data from the SQL server and seamlessly integrates it into the Azure DT model.

\section{Implementing Cyber Attacks at the Edge }
\label{sec:cyberattacks}

In this section, we initiate various cyber attacks at the virtual layer and its physical counterparts on our implemented test bed. We utilize Kali Linux as our attacker box to execute various network attacks. Additionally, we employ an alpha network adapter to establish the Wi-Fi connection on the Kali Linux machine. 
\subsection{Cyber attacks at the Network}
We conduct a series of cyberattacks on smart farms equipped with various IoT sensors. Specifically, deauthentication, ARP poisoning, SQL injection, and port scanning attacks were performed on the testbed for \textit{Smart Farm A}. Similarly, Evil Twin, fingerprinting, DNS spoofing, and SYN flood attacks were executed on the testbed for \textit{Smart Farm B}. Table \ref{tab:attack-summary}
provides a comprehensive list of all cyberattacks, the tools used, associated vulnerabilities, and their impact on other smart farms.
\subsection{Executing Attacks on Digital Twin at the Edge}
\begin{figure}
    \centering
    \includegraphics[width=\columnwidth]{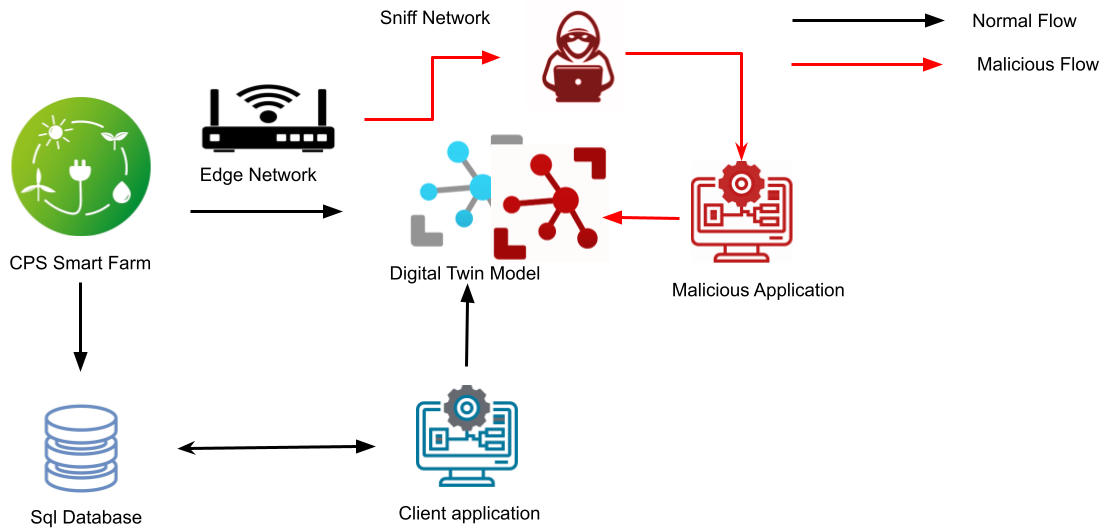}
    \caption{Attacks in Digital Twin}
    \label{fig:DT-attack}
\end{figure}
There are various possible attacks~\cite{karaarslan2021digital,alves2019digital} on DT, where the attacker defeats the DTs' security by manipulating their behaviour or exploiting the update from the physical system to the DTs to steer the CSF into an insecure state. The model corruption technique is employed to attack DT. The fundamental principle of a DT relies on maintaining synchronization between its physical and digital counterparts. If an attacker disrupts this synchronization, the DT model will deviate from its intended behaviour. For example, the attacker may sniff the network gain access to the DT environment, inspect the code, and inject malicious code to corrupt the model. To accomplish this, we develop a malicious Python application that generates simulated data vastly different from the physical counterpart (Fig.~\ref{fig:DT-attack}). The digital model becomes compromised, impacting the accuracy of the DT representation. As a result, the DT may no longer faithfully mirror its physical counterpart, leading to inconsistencies in output.
\subsection{Network Dataset Generation}
Data collection occurred in two phases: (a) capturing network data under normal conditions (b) capturing data during various cyberattacks for each smart farm.

In the first phase, raw IEEE 802.11 frames were captured. Wireshark was then used to monitor WiFi traffic across all channels, identify access points, and list their basic service set identifiers (BSSIDs). If a BSSID was known, filter options were applied to display only the traffic associated with that specific access point. During this data collection phase, Wireshark also gathered all network traffic, which was saved into PCAP files. These PCAP files were stored on the edge servers of \textit{Smart Farm A} and \textit{Smart Farm B}.

The raw dataset was stored as PCAP files and manually parsed for feature extraction. Each PCAP file was opened in Wireshark, filtered to include only the relevant BSSID data, and then saved as CSV files. These CSV files for each smart farm were combined into a single file, with two additional columns added: one indicating whether an attack occurred (class label 0 or 1) and another specifying the type of attack. This same process was followed for \textit{Smart Farm A} and \textit{Smart Farm B}, and the resulting datasets were named Dataset 1 and Dataset 2 respectively. 


\section{Proposed Edge Based Network Anomaly Detection Architecture}
\label{sec:networkarchitecture}
\begin{figure*}[!t]
\centering
\includegraphics[width=0.9\textwidth]{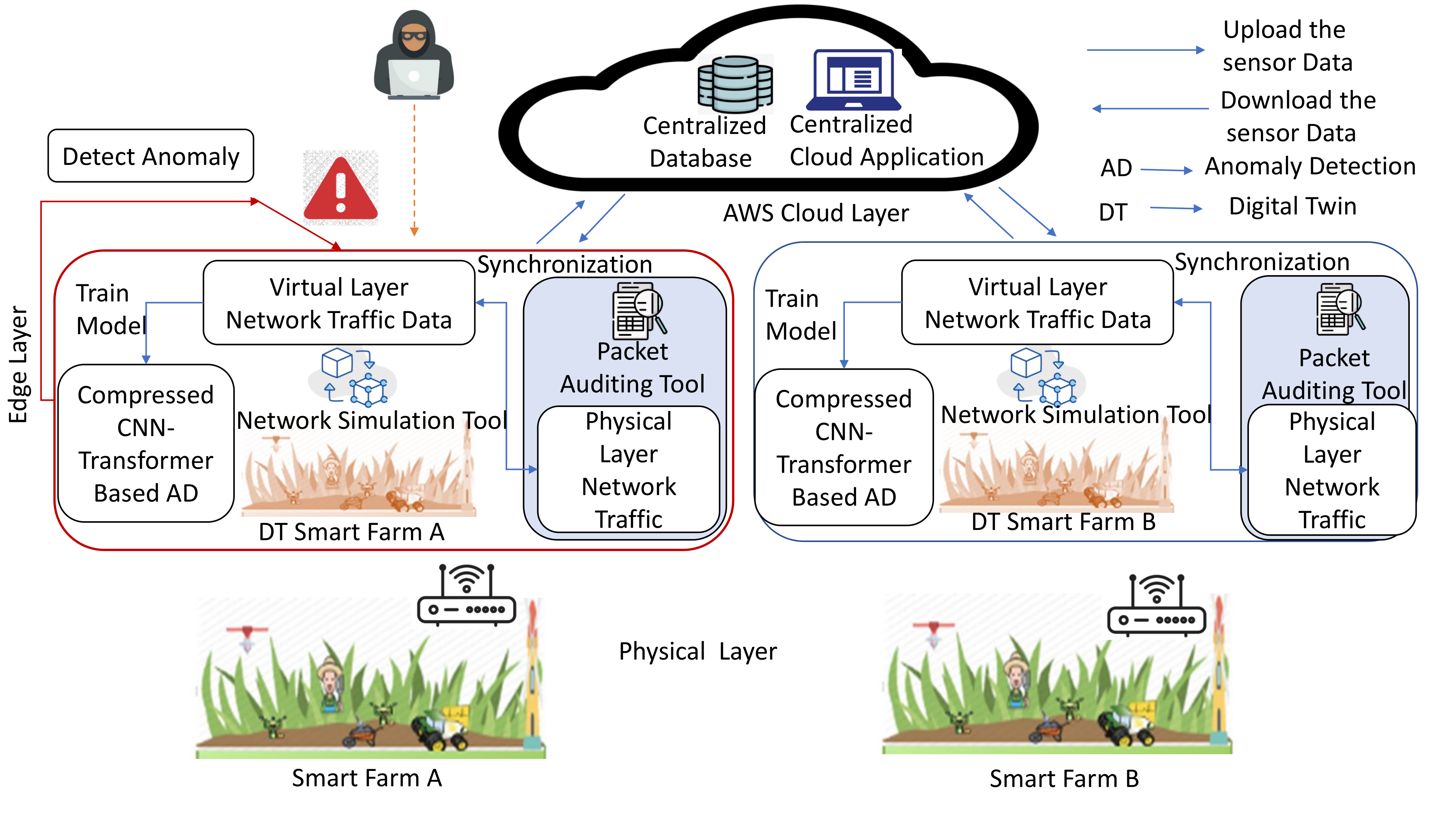}
\caption{Proposed Network Anomaly Detection Architecture for Cooperative Smart Farming}
\label{fig:networkarchtecture}
\end{figure*}
To effectively detect and prevent cyberattacks in CSF, we propose an edge-based network anomaly detection technique that isolates the affected smart farms from the cooperative network upon detecting a cyberattack (Fig.~\ref{fig:networkarchtecture}). The proposed architecture showcases a CSF system in which multiple farms have signed a cooperative agreement. We focus on two smart farms, \textit{Smart Farm A} and \textit{Smart Farm B}, each utilizing distinct smart farming technologies and practices, incorporating unique features, capabilities, and innovations. These solutions may include advanced sensors, communication systems, and intelligent algorithms aimed at optimizing agricultural operations to achieve the overarching goals of smart farming. Each smart farm requires specific computing capabilities, leading to the deployment of a localized computing resource known as the edge server. The edge server facilitates efficient data processing and anomaly detection within each farm. Additionally, each farm can host and deploy DT models and a network simulation tool (Fig. \ref{fig:networkarchtecture}). The network simulation tool models the network topology, traffic load, and the flow of benign and malicious traffic within the network.

The data collected from the DT and network simulation tool can detect malicious activity without interacting with the physical layer, offering predictive insights into network behaviour. This assists in real-time simulation and analysis. In case a physical sensor is compromised, the network simulation can still provide the necessary data to support the anomaly detection model, ensuring continuous monitoring and protection of the smart farm’s network. A packet auditing tool is software utilized to ensure synchronization between simulated network data and real network data. This tool verifies that data packets have not been altered or modified during transmission. Once verified, the network packets are processed by the CNN-Transformer model, which is deployed at each edge layer. If any anomalies are detected, indicating a potential network attack, an alert is sent to the respective smart farm, prompting it to transmit sensor data for further analysis (Fig. \ref{fig:networkarchtecture}). This process ensures that corrupted smart farm is isolated, preventing it from affecting other smart farms within CSF system.

\subsection{CNN-Transformer model}
\begin{figure}[!t]
\centering
\includegraphics[width=\columnwidth]{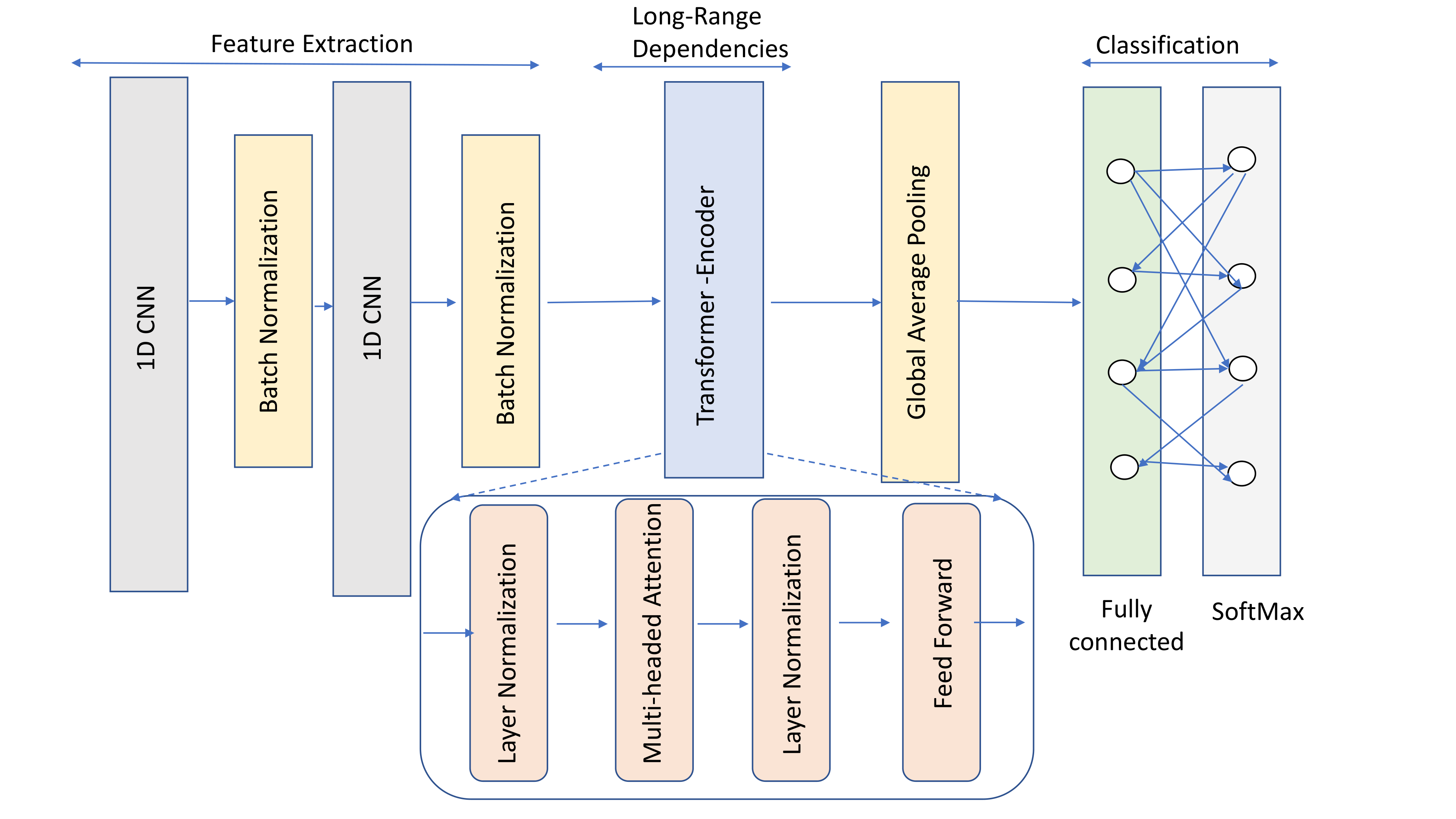}  
\caption{CNN-Transformer Model}
\label{fig:cnntransformer}
\end{figure}

The proposed model integrates a CNN and a Transformer encoder block to efficiently analyze and classify network data (Fig. ~\ref{fig:cnntransformer}). The CNN learns local features from the input data, while the transformer captures relationships in sequential data. A fully connected layer is then used for the final classification.

In the first stage, given a network dataset, the CNN extracts local features. The CNN block consists of 1D convolutional layers (1D CNN) and batch normalization (BN) layers. The convolutional layers detect patterns and features in the network data, which are combined to gain a more comprehensive understanding of network behavior. The batch normalization layers play three key roles: accelerating convergence, preventing gradient issues (explosion/vanishing), and reducing overfitting (Fig.~\ref{fig:cnntransformer}). Transformers, initially designed for NLP, excel at capturing long-range dependencies and relationships in sequential data through their self-attention mechanism. This is especially valuable in network anomaly detection, where correlations across distant time steps or features can be crucial \cite{liu2023intrusion, wu2022rtids}. Layer normalization ensures stability when processing network data of varying lengths by normalizing input embeddings. The multi-head attention mechanism, a core feature of the transformer, allows the model to focus on different parts of the input simultaneously by applying attention multiple times and aggregating the results.

After the attention mechanism, the output passes through a feed-forward layer consisting of a normalized layer and two dense layers. This is followed by a global average pooling layer and a fully connected layer. Finally, a softmax layer is applied to effectively classify different types of network attacks (Fig.~\ref{fig:cnntransformer}).

\subsection{Model Compression using Post Quantization}
While the CNN-Transformer model generalizes the network attacks efficiently, it requires substantial storage space. To effectively deploy the model at the edge, it is essential to compress the model~\cite{gupta2020security,sha2017edgesec}. Commonly used model compression techniques include Quantization, Pruning, and Knowledge distillation~\cite{lu2023hierarchical}. We use the Quantization technique as it does not require altering the model architecture or changing the number of parameters~\cite{sharmila2023qae,jena2024unified}. It simply reduces the numerical precision of the model’s weights and activations. Specifically, we use the Post-Quantization technique, where quantization is applied after the model training. It converts the model's parameters (weights) from 32-bit floating-point numbers to a more efficient, lower precision (like 8-bit integers)~\cite{lu2023hierarchical,jena2024unified}. 

\section{Performance Analysis}
\label{sec:analysis}
This section involves the pre-processing of the dataset, model training, and performance analysis and evaluation of our proposed model.
\subsection{Data pre-processing}
The pre-processing step involves one hot encoding, label encoding, normalization, and standardization of network datasets. We first convert the categorical and string values into integer values in the dataset. The labels are in the form of a string, so we convert them to integers. Standardization was applied to all the captured features by calculating their standard deviations to ensure consistency across the data. We also removed certain features because they were unnecessary for the anomaly detection system. Features like `Frame number', `Source IP', `Destination IP', `Timestamp', `Source port', and `Destination port' were excluded to prevent the models from simply learning patterns based on numbering on specific IP addresses and ports, which could lead to overfitting or biased predictions. Then, the dataset is divided into training, validation, and testing sets. 
\subsection{Model Training}
After pre-processing, the compiled model is trained on the pre-processed training data using the model. fit() function. It employs cross-entropy as the loss function and the Adam optimizer for optimization. During training, the weights are updated based on the computed gradients and learning rate to enhance the model's predictions in subsequent forward passes. After each epoch, the model's performance is evaluated on the training and validation sets. This process is repeated over  10 to 50 epochs, allowing the model to improve its performance progressively. The learning rate is set to 0.001, and the batch size is 32. The training and evaluation are conducted on two edge devices using Google Colab, equipped with the same hardware specifications: 18 GB of RAM and 108 GB of hardware. Finally, the test dataset evaluates the model's performance after training.
\subsection{Model Evaluation}
Various performance matrices, such as accuracy and F1 score, were used to validate the effectiveness of our proposed model. 
The accuracy of a model is defined as:
\[
\text{Accuracy} = \frac{\text{True Positives} + \text{True Negatives}}{\text{Total Number of Samples}}
\]

The F1 score is the mean of precision and recall, defined as:
\[
\text{F1 Score} = 2 \times \frac{\text{Precision} \times \text{Recall}}{\text{Precision} + \text{Recall}}
\]
Where,
\[
\text{Precision} = \frac{\text{True Positives}}{\text{True Positives} + \text{False Positives}}
\]
\[
\text{Recall} = \frac{\text{True Positives}}{\text{True Positives} + \text{False Negatives}}
\]
True Positive represents the number of attack samples correctly classified, and False Positive represents the number of benign samples incorrectly classified. True negative is the number of benign samples correctly classified, and False Negative represents the number of attack samples incorrectly classified.
\begin{figure}[!t]
\centering
\includegraphics[width=0.9\columnwidth]{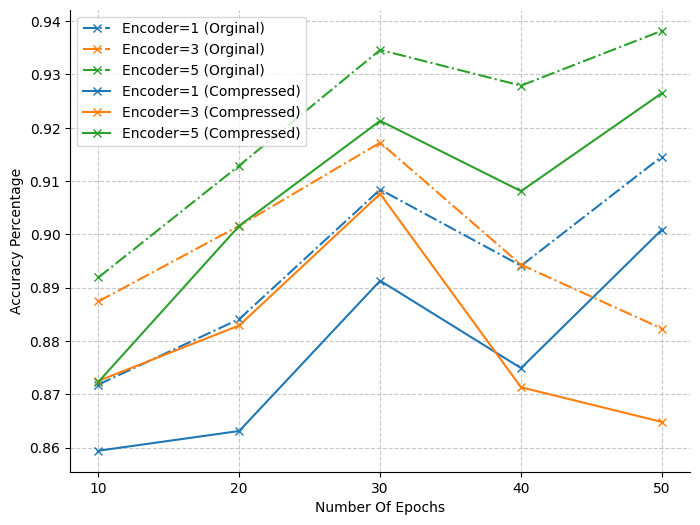}  
\caption{Comparison  Accuracy In Smart Farm A when Encoder=1,3,5 (Original Vs Compressed)}
\label{fig:cnntransformersmartfarm1}
\end{figure}

\begin{figure}[!t]
\centering
\includegraphics[width=0.9\columnwidth]{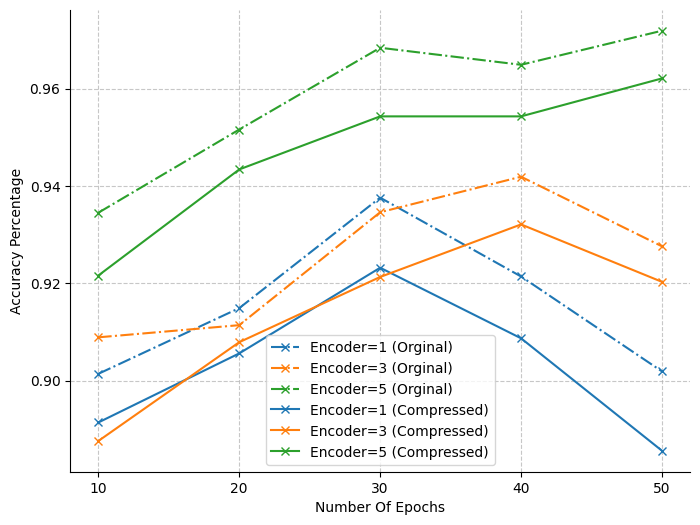}  
\caption{Comparison  Accuracy In Smart Farm B when Encoder=1,3,5 (Original Vs Compressed)}
\label{fig:cnntransformersmartfarm2}
\end{figure}
 We evaluate the performance of the CNN-Transformer models using two datasets, Dataset 1 and Dataset 2, for two edges, \textit{Smart Farm A}  and \textit{Smart Farm B}, respectively (Fig.~\ref{fig:cnntransformersmartfarm1}, \ref{fig:cnntransformersmartfarm2}). We also assess the impact of varying encoder layers. Initially, the model was trained and validated with a single encoder layer, and the number of encoder layers was gradually increased to 3 and 5 while keeping the CNN layer configuration constant. As shown in Fig.~\ref{fig:cnntransformersmartfarm1}, increasing the number of encoder layers improved model accuracy. 
For example, with a single encoder layer and 10 epochs in \textit{Smart Farm A}, the model achieved an accuracy of 87\%, which gradually increased to 90\% after 30 epochs and reached 91\% at 50 epochs. When using three encoder layers, the accuracy of \textit{Smart Farm A} slightly improved to 91\% at 30 epochs but gradually declined as the number of epochs increased. However, with five encoder layers, the model's performance significantly improved, achieving around 94\% accuracy at 50 epochs. This demonstrates that increasing the number of transformer layers enhances the model's ability to detect attacks.

As shown in Fig.~\ref{fig:cnntransformersmartfarm2}, the CNN-Transformer model performs better in \textit{Smart Farm B} than in \textit{Smart Farm A}. For example, with a single encoder layer, the model achieved its best accuracy of 93\% at 30 epochs, which increased to 94\% at 40 epochs. With five encoder layers, the model's performance further improved, reaching 97\% accuracy at 50 epochs. This indicates that similar to \textit{Smart Farm A}, increasing the number of encoder layers in \textit{Smart Farm B} enhances the model's ability to detect network attacks efficiently.

\begin{figure}[!t]
\centering
\includegraphics[width=0.9\columnwidth]{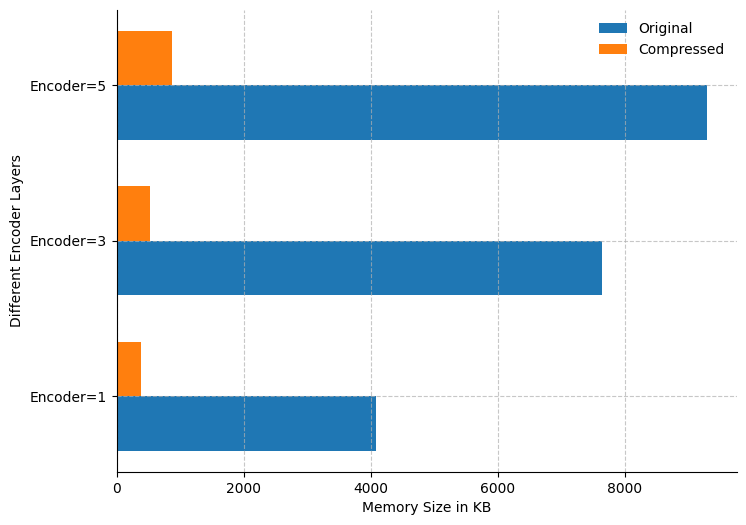}  
\caption{Comparison of Memory Size in KB of trained CNN-Transformer model 50 number of epochs with different Encoder Layers In Smart Farm B}
\label{fig:memorysize}
\end{figure}

After model training, we compress the model using Post-Quantization to deploy the model at the edge effectively. As illustrated in Fig.~\ref{fig:cnntransformersmartfarm1}
and \ref{fig:cnntransformersmartfarm2}
, the solid line represents the accuracy of the model after compression. It can be observed that model compression leads to a slight drop in accuracy, approximately 1-2\%. For instance, in \textit{Smart Farm A}, with 50 epochs and five encoder layers, the model's accuracy decreases from 93\% to 92\% post-compression. Similarly, in \textit{Smart Farm B}, the accuracy drops from 97\% to 96\% under the same conditions. However, the primary advantage of model compression is the significant reduction in memory usage. As shown in Fig.~\ref{fig:memorysize}, the memory size is considerably reduced across all encoding layers after model compression in \textit{Smart Farm B}. For example, with one encoder layer, the memory size decreases from 4081.62 KB to 383.76 KB, achieving a compression ratio of 90.60\%.
\subsection{Comparison with the compressed CNN-Transformer model vs Traditional ML models}
In the next set of experiments, we compare the compressed CNN-Transformer model with the five encode layer's performance with the other ML models such as Random Forest (RF), Logistic Regression (LR),
CNN and Long Short-Term Memory (LSTM) models. The detection accuracy and F1 score of the compressed CNN-Transformer model Vs Traditional ML Models in \textit{Smart Farm A} are shown in Fig.~\ref{fig:comparisionsmartfarm1}, whereas the accuracy and F1
score of the compressed CNN-Transformer model vs. traditional ML Models in \textit{Smart Farm B} are shown in Fig.~\ref{fig:comparisionsmartfarm2}. The Fig.~\ref{fig:comparisionsmartfarm1} and \ref{fig:comparisionsmartfarm2} show that the compressed CNN-Transformer model achieves higher accuracy and F1 scores on both the \textit{Smart Farm A} and \textit{Smart Farm B}. For \textit{Smart Farm A}, the accuracy of the compressed CNN-Transformer model is higher than that of the CNN-LSTM model, with 90\% compared to 92\%. Similarly, on \textit{Smart Farm B}, the compressed CNN-Transformer with five encoder layers outperforms the CNN-LSTM, achieving 3\% higher accuracy and F1 score. This shows that applying model compression to a CNN-Transformer model can still better generalise network classification tasks compared to traditional ML models.
\begin{figure}[!t]
\centering
\includegraphics[width=0.9\columnwidth]{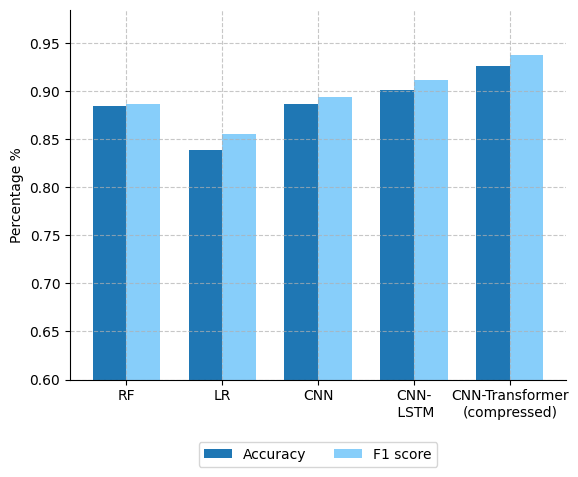}  
\caption{Comparison of Accuracy and F1 score of Compressed CNN-Transformer (5 Encoder) with different ML Model In Smart Farm A}
\label{fig:comparisionsmartfarm1}
\end{figure}

\begin{figure}[!t]
\centering
\includegraphics[width=0.9\columnwidth]{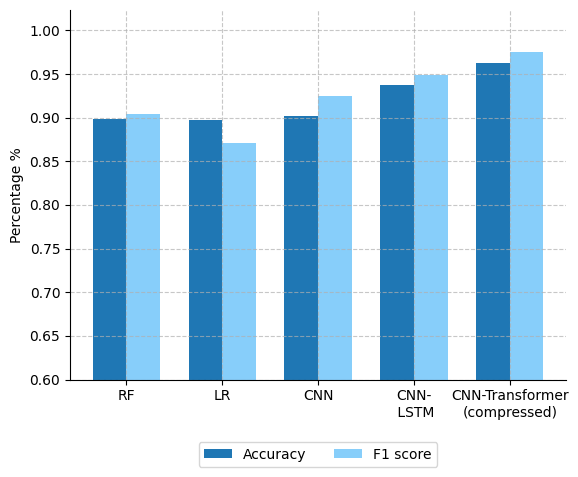}  
\caption{Comparison of Accuracy and F1 score of Compressed CNN-Transformer (5 Encoder) with different ML Model In Smart Farm B}
\label{fig:comparisionsmartfarm2}
\end{figure}

\section{Conclusion and Future Work}

\label{sec:conclusion}
CSF is emerging as a vital component of critical infrastructure and has an efficient role in contributing to a country's GDP. To protect this vital sector, this research investigates the impact of cyberattacks on the physical and DT layers in CSF and proposes a CNN-Transformer-based network anomaly detection model to detect and mitigate it. Initially, we explore the challenges inherent in CSF and briefly outline the involved layers through Edge-enabled architecture. We establish a test bed for CSF and conduct various network attacks on each edge. Additionally, we implement the DT at the edge using Azure DT and subject it to attacks. We generated two smart farming network datasets at the edge, named Dataset 1 and Dataset 2. To detect and prevent cyberattacks at the DT and its physical counterpart, we propose a CNN-Transformer-based network anomaly detection model at the edge and evaluate its effectiveness in our two smart farming network datasets using varying numbers of encoder layers. Our results indicate that the model's ability to detect the network attack increases with the increasing number of encoder layers. However, this enhancement comes at the price of higher memory, leading to a significant challenge for edge deployment. To address this, we compress the model using Post-Quantization techniques. This compression method effectively diminishes the model's memory size with minimum impact on accuracy. We also compared our compressed model's performance with traditional machine learning approaches, and the results showed that the compressed CNN-Transformer model outperformed them in both accuracy and F1 score. 

In future work, we want to expand our research by developing a collaborative network anomaly detection model capable of generalizing across both smart farming datasets including DT datasets. This will enable the model to detect cyberattacks on other smart farms within the CSF network effectively. By addressing these aspects, we aim to enhance the security of CSF systems against emerging cybersecurity challenges, ultimately enhancing the protection of agricultural operations.

\section*{Acknowledgment}
This work is partially supported by the US National Science Foundation grants 2230609,  2226612 and 2346001.

\bibliographystyle{IEEEtran}
\bibliography{citation.bib}

\end{document}